\documentclass[preprint,aps,nofootinbib,showpacs]{revtex4}
\usepackage{graphics}
\usepackage{epsfig}

\draft

\begin{document}

\title{On the absorbing-state phase transition in the one-dimensional
triplet creation model}

\author{G\'eza \'Odor$^\dagger$ and Ronald Dickman$^\ddagger$}

\affiliation{
$^\dagger$MTA-MFA Research Institute for Technical Physics and Materials Science\\
$^\ddagger$Departamento de Fisica and National Institute of Science and
Technology of Complex Systems, ICEx,
Universidade Federal de Minas Gerais, Caixa Postal 702,
30161-970, Belo Horizonte - Minas Gerais, Brasil
}

\begin{abstract}
We study the lattice reaction diffusion model $3A\to 4A$,
$A\to\emptyset$ (``triplet creation") using
numerical simulations and $n$-site approximations.
The simulation results suggest that the phase transition is discontinuous
at high diffusion rates.  In this regime the
order parameter appears to be a discontinuous function of the creation rate;
no evidence of a stable interface between active and absorbing phases is found.
Based on an effective mapping to a modified compact directed percolation process,
we shall nevertheless argue that the transition is {\it continuous}, despite
the seemingly discontinuous phase
transition suggested by studies of finite systems.

\end{abstract}
\maketitle

\section{Introduction}

The exploration of phase transitions in simple, one component
nonequilibrium models has attracted considerable
interest \cite{DickMar,Hayeof,Orev,Svenrev}, and important steps
towards identifying the related universality classes have been achieved \cite{EK06,Obook08}.
In nonequilibrium models phase transitions may occur even in one-dimensional
systems:  the well known arguments, due to Landau and to van Hove
\cite{Hove}, against
phase transitions in one-dimensional systems with short-range
interactions, do not apply in the absence of detailed balance.
However in low dimensions the effect of fluctuations is stronger,
making continuous phase transitions more common.  (A familiar example
is the three-state Potts model, which exhibits a continuous transition
in two dimensions, and a discontinuous one for $d > 2$.)

In one dimension, discontinuous phase transitions have been found in
models with long-range interactions \cite{boccikk}, or a
conserved density \cite{gla63,cpccikk}, and in multi-component systems
\cite{EKKM98,GLEMAA95,woh,MD07}.
Compact directed percolation (CDP) has a discontinuous transition between
a pair of absorbing states (all sites full or all empty) \cite{domanykinzel,Essam}; a
similar transition between absorbing states is found in the
one-dimensional Ziff-Gulari-Barshad model \cite{ZGB}.
A discontinuous transition between and active phase
and an absorbing one in a single-component model was claimed for
the triplet creation model (TCM) \cite{Dick1st}, which does not possess a
conservation law or long-range interactions.
This model features particle reactions $3A\to 4A$, $A\to\emptyset$ and
explicit diffusion (hopping) \cite{explicit}.
On increasing the diffusion probability,
a crossover from a continuous to a discontinuous phase transition was
detected in simulations and cluster mean-field approximations.
Similar behavior was also reported in a stochastic cellular automaton \cite{boccikk}.

Subsequently, Hinrichsen argued that in one dimension, discontinuous transitions
between an active and an absorbing state cannot exist in models like the
TCM \cite{Hayefirsto,Hbook}.
The original findings for the TCM \cite{Dick1st} were nevertheless
confirmed in
spreading simulations by Cardoso and Fontanari \cite{CF06-1st} and in
fixed order-parameter
simulations by Fiore and de Oliveira \cite{fiore}.
The spreading exponents are shown in \cite{CF06-1st}
to be those of compact directed percolation (CDP) \cite{domanykinzel,Essam};
a tricritical point is suggested
for a diffusion probability $D \simeq 0.95$.
Very recently Park \cite{park09} reported simulation results that again support a continuous
phase transition, belonging to the directed percolation (DP) universality class,
at high diffusion rates.

Recently, a field theoretic analysis of bosonic reaction-diffusion (RD) models
led to a hypothesis \cite{EK06}, based on a general phase transition
classification scheme: {\it bosonic, one-component RD
systems with $n$-particle creation and $m$-particle annihilation
always exhibit a first-order transition if $n>m$}.
This is indeed the case above the upper critical dimension
(see \cite{tripcikk}). However in bosonic models one has
to introduce a  higher-order coagulation term
$m'A\to (m'-l)A$ with ($m'>n$), to avoid an infinite particle
density in the active phase.
Furthermore the topological phase space method used in \cite{EK06}
deals with the reactions (creation and annihilation),
but does not take into account the effect of diffusion,
which turns out to be relevant in some cases,
when different reactions compete \cite{Od23402d,Od1240,CCD04}.

In this work we study the TCM in an effort to determine
whether multi-particle creation, combined with rapid diffusion, can
overcome fluctuations and generate a discontinuous phase transition in one dimension.
This is a problem of longstanding interest in nonequilibrium statistical physics,
and is related to the existence of first-order
depinning transition in nonequilibrium wetting
(i.e., in a system with multiplicative noise, with an attractive wall)
\cite{Mrev04}.

The remainder of this paper is organized as follows.  In Sec. II we
define the model, and review applicable simulation methods and previous
results regarding the nature of the phase transition.
Section III is devoted to a discussion of $n$-site approximations, and
Sec. IV to our simulation results.  In Sec. V we use these results to motivate
a simplified description of the model in the high diffusion rate regime,
and discuss the nature of the transition using this mapping.
Finally in Sec. VI we summarize our findings.

\section{The triplet creation model}

The TCM is defined on a lattice, with each site either vacant or occupied
by a particle; multiple occupancy is forbidden \cite{Dick1st}.
In the one-dimensional TCM,
a particle ($A$) attempts diffusion at rate $D \leq 1$, creation ($3A \to 4A$) at rate
$\lambda (1-D)/(1+\lambda)$, and is annihilated ($A \to 0$) at rate
$(1-D)/(1+\lambda)$.
In a diffusion attempt, one of the nearest neighbor (NN) sites of the particle is
chosen at random, and the particle jumps to this site if it is empty.
If the target site is occupied, the configuration remains the same.
In a creation attempt, if both NN sites of the particle are occupied, then
one of the second-neighbor sites of the central particle is chosen at random,
and if this site is empty, a new particle is placed there.
If the conditions of two occupied NN sites and an empty target site
are not fulfilled, the configuration does not change.
Annihilation occurs independently of the states of neighboring sites.
The configuration with all sites empty is absorbing.  Since the sum of these transition
rates is unity, the total transition rate in a system with $N$ particles is simply $N$.
In simulations, the time increment associated with each attempted event
(whether accepted or not) is $\Delta t = 1/N$, and one Monte Carlo step (MCS)
corresponds to an accumulated time increment of unity.

In Ref. \cite{Dick1st} the one-dimensional TCM was shown to exhibit a phase transition between
the active and absorbing states; the transition was found to be continuous
(and in the DP universality class) for smaller diffusion
rates, but discontinuous for large $D$.
By a discontinuous transition we mean one in which
the order parameter is a discontinuous function of the relevant
control parameter(s), in the infinite-size limit.
In the TCM the order parameter is the particle density
$\rho$, and the control parameters are $\lambda$ and $D$.  Since one of the phases
is absorbing, at a discontinuous transition $\rho$ should jump between zero and
a finite value.

The characterization of a transition as continuous or discontinuous
in numerical simulations is fraught with difficulties:
finite-size rounding can mask the discontinuity, and
any finite system must eventually become trapped in the absorbing state.
To circumvent these problems,
a number of strategies have been proposed.

{\it Hysteresis with a weak source.}
A characteristic feature of
discontinuous phase transitions is hysteresis.
If one of the phases is absorbing, however, hysteresis cannot be
observed simply by varying a control parameter, since
the absorbing phase allows no escape.
Bideaux, Boccara and Chat\'e \cite{Bid} showed that
when the transition is discontinuous, adding a weak source of
activity changes the absorbing and active
phases to {\it low-activity} and {\it high-activity} phases, respectively.
One may then observe a hysteresis loop between these phases, on varying the
control parameter.
This approach was used in \cite{Bid} to demonstrate a discontinuous
phase transition in a probabilistic cellular automaton,
and was applied to the TCM in \cite{Dick1st}, yielding a hysteresis loop.
Below, we shall revisit the question of scaling under a weak source.

{\it Conserved order parameter simulations.}
In conserved order parameter simulations \cite{conscp},
particles are neither created nor destroyed.  Changes in configuration
occur through particle jumps, which can be of any size up to that of the
entire system, in a manner that respects the local rules of the process.
The simulation yields an
estimate for the control parameter value corresponding to the chosen order
parameter density.
Using this method, Fiore and de Oliveira found
evidence for a discontinuous transition in both the TCM and the related pair
creation model (with creation reaction $2A \to 3A$) at high diffusion rates \cite{fiore}.

{\it Quasistationary (QS) simulation.}
As in conserved order parameter simulations, QS simulation removes the absorbing state from the
dynamics, but in a manner that samples the quasistationary probability distribution
(i.e., conditioned on survival) \cite{qssim}.
A study of the TCM using this method \cite{MD07} showed that as the system size
tends to infinity, the QS order parameter appears
to develop a discontinuity between zero and a positive value, as $\lambda$ is varied
at a high diffusion rate, $D=0.98$.  (For a finite system the discontinuity is of
course rounded.)  A study of
the TCM with biased diffusion (hopping in one direction only) yielded evidence of a
sharp discontinuity \cite{rdunp}.

{\it Spreading simulations.}
Studies of the spread of activity, starting from a seed at the origin, have long been employed
to characterize continuous phase transitions to an absorbing state \cite{torre,DickMar}.
At the critical point, the survival probability $P(t)$, mean number of active sites $n(t)$,
and mean-square distance $R^2(t)$ of active sites from the origin, all follow power laws.
At a discontinuous transition, there is in principle no reason to expect scale-invariant
spreading dynamics.  Nevertheless, in the case of the TCM, Cardoso and Fontanari \cite{CF06-1st} demonstrated
power-law spreading at the transition point, $\lambda_c$, for $D=0.98$.  The scaling exponents were identified
as those of CDP, which, as noted above, suffers a discontinuous transition between a pair of
symmetric absorbing states.

{\it Interface motion.}
Suppose we prepare the system with all sites occupied, allow it to relax to the QS state,
and then remove all particles from half of the lattice.  In the subsequent evolution,
the interface between active and inactive regions broadens due to diffusion, and in general
drifts toward one region or the other.  Below we report studies showing
that the drift velocity is proportional to $\lambda - \lambda_c$.
In a related analysis, we initialize the system with all sites in the region $1,...,M$ occupied,
and sites $M+1,..,L$ empty, and study the long-time survival probability $P(M)$.  At the
transition, the dependence of $P(M)$ on $M$ is consistent with independent, randomly diffusing interfaces,
as in CDP.  This result supports the existence of two phases, one absorbing, the other active,
separated by a large gap in density.  The two phases do not coexist: the fluctuating interfaces
eventually meet, and one of the phases is lost from the system.

Summarizing, the above mentioned studies, some from the recent literature, others to be
reported below, provide evidence for a discontinuity in the QS order parameter,
for hysteresis, and for a connection between the TCM at high diffusion rate and compact directed
percolation.

\subsection{Hinrichsen's objection}

Some years ago, Hinrichsen presented an argument to the effect that discontinuous phase transitions
between an active and an absorbing state are impossible in one-dimensional systems with local
interactions, and without additional conservation laws, special boundary conditions,
or macroscopic currents \cite{Hayefirsto,Hbook}.
The argument is based on the observation that the
effective surface tension of interfaces in such systems does not
depend on the size of the domains they delimit.
Hinrichsen's argument prohibits the presence of fixed, stable boundaries between
coexisting phases; as noted, no such boundaries have been observed in simulations.
But this in itself does not
appear to imply that the dependence of the order parameter on growth rate must be continuous
at the transition. The one-dimensional totally asymmetric exclusion process (TASEP), for
example, exhibits a discontinuous phase transition in a certain region of parameter space,
even though the position of the boundary between high- and low-density phases fluctuates
over the entire system \cite{DEHP92,SchDom93}.

Hinrichsen \cite{Hayefirsto} also reported simulation results supporting
a continuous transition in the TCM at diffusion rate $D=0.9$, that is, above the
estimate for $D_t$ given in \cite{Dick1st}.  It is now generally acknowledged that
$D_t > 0.9$ in the TCM.

In a recent study \cite{park09}, Park reported simulation results that support DP-like scaling
in the TCM at diffusion rates 0.95 and 0.98.  Specifically, the
order parameter (starting from a filled lattice) appears to decay at long times as
$\rho(t) \sim t^{-\delta}$, with $\delta$ taking its DP value, over about two decades
in time.  We note however that the decay exponent is very sensitive to the choice of the
time interval used for analysis and of the control parameter $\lambda$.
Analyzing simulation results for $D=0.98$ in studies extending to $10^9$ MCS, we obtain
local decay exponents $\delta_{eff}$ between 0.1 and 0.2, varying $\lambda$
in a very narrow range.
A crossover between
a long supercritical plateau for $t \le 10^7$ and a rapid decay to an
inactive state cannot be ruled out.
For $D=0.98$, the regime during which DP-like scaling is found in Ref. \cite{park09}
(i.e., $10^7 \leq t \leq 10^9$) corresponds to overall particle densities in the range
0.67 - 0.32.  While scaling behavior can be observed at such densities in the contact
process \cite{DickMar}, definitive results for the decay exponent would require studying systems with
substantially smaller values of $\rho$.
Finally, three or more critical exponents would have to be determined to demonstrate
convincingly that the transition falls in the DP class.

Although the results of Hinrichsen and of Park do not appear to rule out rigorously a discontinuous
transition in the TCM, we believe that they are fundamentally correct.  This conclusion is based
not on simulation results but rather on
a mapping to a modified CDP process, to be
developed in Sec.~V.  Analysis of this mapping in leads to the conclusion that the transition is in fact
continuous, despite abundant numerical evidence to the contrary.

\section{$n$-site approximations}

One of the most common theoretical approaches to
Markov processes with spatial structure
is a truncation of the master equation known as an $n$-site approximation \cite{avraham}.
Such approximations have been applied to the TCM in efforts to
determine the order of the transition; in this section we review and extend these results.

The simplest method in this family
is dynamic mean-field theory or the {\it one-site approximation}, in which the
probability of an $m$-site configuration is factored into a product of $m$ single-site
probabilities, so that, for example, $P(\bullet \bullet \bullet \, \circ) \simeq \rho^3 (1-\rho)$
where $\bullet$ ($\circ$) denotes an occupied (vacant) site and $\rho$ is the fraction
of occupied sites.  The resulting equation for $d \rho/dt$ yields rather poor predictions for
the TCM; better results are obtained using larger clusters.
In the $n$-site approximation, the equations that govern the probability distribution for
clusters of $n$ sites are truncated by expressing the probabilities of $n+1$ site
(or larger) clusters in terms of the $n$-site distribution.
In the 3-site approximation, for example,
we write $P(\bullet \bullet \bullet \, \circ) \simeq
P(\bullet \bullet \bullet) P(\bullet \bullet \circ)/ P(\bullet \, \bullet)$. As $n$ grows, the number
and complexity of the equations increases rapidly, but it is possible to generate the equations,
and integrate them numerically, via a computational algorithm \cite{mancam}.

In Ref. \cite{Dick1st}, the 4-site approximation for the TCM was found to predict a continuous
phase transition for diffusion rates $D<D_t$ and discontinuous one for $D>D_t$.  The
predicted value for the tricritical diffusion rate $D_t$, however, is much smaller
than that reported in simulations ($D_t \simeq 0.95$).  Since the phase diagram
predicted by the $n$-site approximation generally converges to the correct one as $n \to \infty$,
it is of interest to study the results for larger $n$.
In certain cases, predictions based on a sequence of
$n$-site approximations behave in a consistent manner, and can be extrapolated
to provide estimates of the transition point and critical exponents, via the
coherent anomaly method \cite{katori,mancam}.

\begin{figure}[ht]
\begin{center}
\epsfxsize=70mm
\epsffile{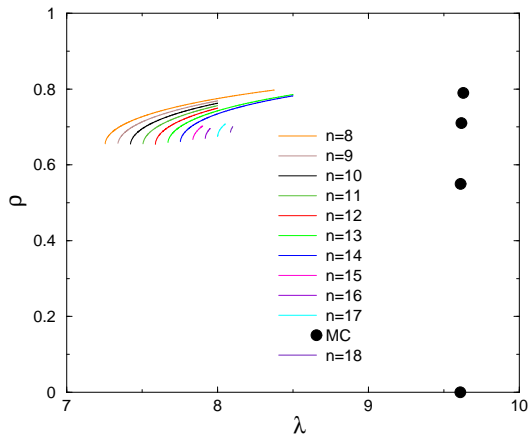}
\caption{(Color online) Order parameter versus creation rate
in the one-dimensional TCM with $D=0.98$, in the $n$-site approximation
with $n=8,...,17$; points:
simulation.}
\label{cs_98}
\end{center}
\end{figure}

For small $n$, the position of $D_t$ varies considerably.
For example, the $n \le 3$ approximations
yield a discontinuous transition even for $D=0$, but for
$n \geq 4$ there is a tricritical point at some $D_t > 0$.
The estimates for $D_t$ increase gradually with $n$; for
$n=8$,  one finds $D_t > 0.5$, for example.  On the other hand,
for a fixed, large diffusion rate, the transition remains discontinuous,
with a large jump in the order parameter, which does not diminish
appreciably with increasing $n$. For $D=0.98$, for example, Fig.~\ref{cs_98}
shows that all the approximations studied ($n \leq 17$) yield a discontinuous
transition.

Recently, Ferreira and Fontanari published results casting doubt
on the utility of $n$-site approximations for the TCM \cite{ff09}.
They show, for example, that for $D=0$, the values of $\lambda_c$
veer {\it away} from the simulation value as $n$ is increased from 11 to 18;
our studies confirm this observation.
(We note that a nonmonotonic approach to the critical point is
observed in a
stochastic cellular automaton in which at least three particles are
required for particle generation or survival \cite{boccikk}.)
Moreover, the values of the tricritical reproduction rate $\lambda_t (n)$
(for $n \leq 14$) appear to converge to an unphysical (negative) value as $n \to \infty$ \cite{ff09}.

\begin{figure}[ht]
\begin{center}
\epsfxsize=70mm
\epsffile{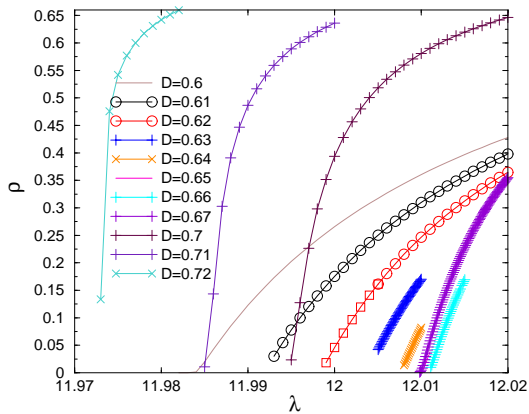}
\caption{(Color online) Order parameter versus creation rate
in the one-dimensional TCM in the $13$-site approximation.}
\label{cs13tr}
\end{center}
\end{figure}

Thus if the $n$-site approximations converge to the correct values, they
do so in a nonmonotonic fashion, such that for the cluster sizes accessible
with present technology ($n \leq 20$ or so), quantitative results for
$\lambda_c$ cannot be obtained for the TCM.
Although our $n$-site approximations show stable nonvanishing gap sizes
at $D=0.98$ for $n \leq 18$ (Fig.~\ref{cs_98}), we find that for a given
level ($n=13$, say), $\lambda_c(D)$ is not a monotonic function of $D$,
and the location of the tricritical point is rather uncertain.
Fig.~\ref{cs13tr} shows that on increasing the diffusion rate from
$D=0.6$ to $D=0.66$, the critical point shifts to higher values, but
for $D>0.66$ this tendency reverses. On the other hand, there is no evidence
of a discontinuous transition for $0.67 \le D \le 0.71$. Our analysis
provides a higher $\lambda_t$ estimate for $n=13$ than found in
\cite{ff09}; the reason for this difference is not known, since our
result for $\lambda_t (n=4)$ agrees with that reported in the latter work.
We observe a similar behavior for $n=14$ and $n=15$; our tricritical
point estimates do not fit on the extrapolation line given in
\cite{ff09}.

For the cluster sizes studied, $\lambda_t(n)$ cannot be fit with a
linear function of $1/n$, so that the $n\to\infty$ limiting value cannot be
estimated with confidence.  It seems likely that $\lambda_t(n)$ exhibits an
oscillatory convergence with $n$; if so, the question of whether
$\lim_{n \to \infty} D_t (n) < 1$,
(that is, the existence of a discontinuous transition), cannot be resolved
using the available $n$-site approximation results.

\section{Simulation results}

We study the TCM via Monte Carlo simulation, using several approaches that complement
earlier analyses: dependence on the initial value of the order parameter, interface dynamics,
scaling in the presence of a weak
source of activity, and scaling of the quasistationary order parameter.
While some of the results would seem to provide good evidence of a discontinuous transition,
we shall defer our conclusion until the following Section.

\subsection{Initial density dependence}

If a phase
transition is discontinuous, the evolution of the system should depend strongly
on the initial condition, while at a continuous transition the evolution
is toward the same QS state, regardless of the initial condition.
In \cite{DM08}, simulations of the TCM at diffusion rate $D=0.98$ are reported,
showing that at the transition ($\lambda_c \simeq 9.60$), the value of the
order parameter at long times depends on its initial value.
For initial particle densities $\rho(0) $ between 0.3 and unity, the system
evolves to the active state, while for
$\rho(0) \leq 0.3$ it rapidly approaches the absorbing state.  (In studies using $\rho(0) < 1$,
the initially occupied sites are chosen at random, uniformly over the lattice.)
These results demonstrate that an active phase, characterized by
a high value of the order parameter, is accessible starting from a high density,
but not from a low one.
The findings for high diffusion rate are in sharp contrast to those found for $D=0$,
for which the critical reproduction rate is $\lambda_c=12.015$.
In this case, the particle density attains the same QS value, starting
from very different initial values.

\begin{figure}[ht]
\begin{center}
\epsfxsize=70mm
\epsffile{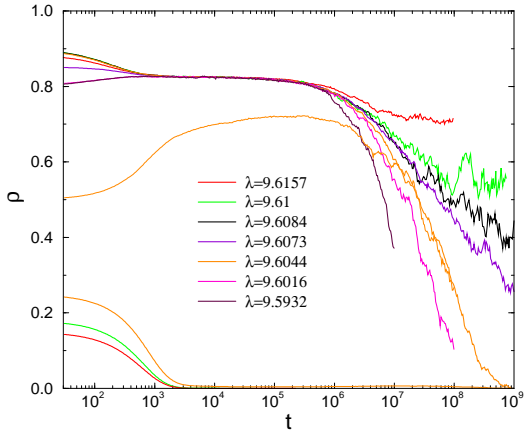}
\caption{(Color online) TCM: $\rho(t)$ for $D=0.98$, using various initial particle
densities, and creation rates near the transition value, as indicated.}
\label{3410_98}
\end{center}
\end{figure}

Here we extend the simulations of \cite{DM08} to
much larger systems.
We follow the evolution of the order parameter $\rho(t)$ in systems of $L=2\times 10^5$ sites (with
periodic boundary conditions) for times of up to $10^9$ MCS, averaging over 5 -20 realizations.
For $D=0.98$, we find that for low initial values
($\rho(0)\le 0.25$), the density falls exponentially
for $\lambda \le 9.616$ (see Fig.\ref{3410_98}).

\begin{figure}[ht]
\begin{center}
\epsfxsize=70mm
\epsffile{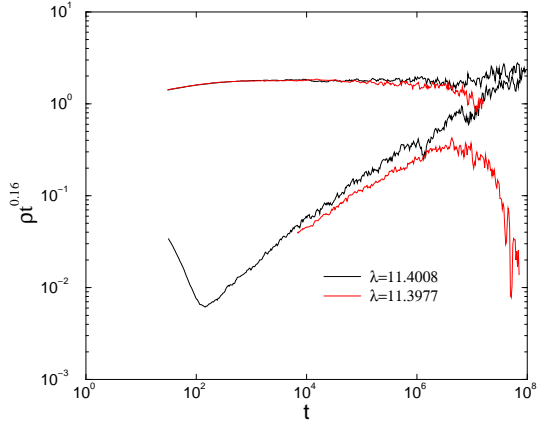}
\caption{(Color online) $\rho(t)$ as in Fig. \ref{3410_5}, but for $D=0.5$}
\label{3410_5}
\end{center}
\end{figure}

For the same reproduction rates, using $\rho(0) > 0.25$,
the order parameter $\rho(t)$ exhibits a long plateau
($10^3<t<10^5$ MCS) at a high density. At longer times $\rho(t)$ decays,
as expected in a finite system.
For $D=0.5$, by contrast, the order parameter curves $\rho(t)$, starting from high and low
initial values, attain a common value at long times
(see Fig.\ref{3410_5}).

\subsection{Fluctuating boundary studies}

\begin{figure}[ht]
\begin{center}
\epsfxsize=150mm
\epsffile{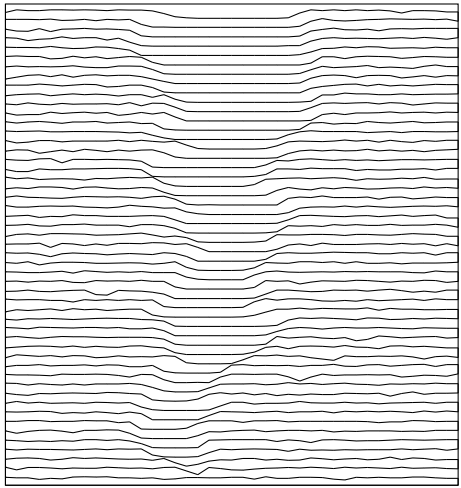}
\caption{Space-time evolution of particle density for $D=0.98$, $\lambda=9.60$, and $L=2000$,
starting from a fully occupied region of 1500 sites and the remainder empty.
Time increases downward, with each sweep (at intervals of 10\,000 time units) showing
the density profile averaged over blocks of 50 sites.}
\label{qcnc7}
\end{center}
\end{figure}

In these studies the initial configuration consists of two blocks,
one fully occupied, of $n_0$ sites, and the other,
of $L-n_0$ sites, completely empty.
For $D=0.98$ and $\lambda=9.60$, one finds that in the initially occupied region,
the particle density quickly relaxes to its QS
value of about $\rho_{QS} \simeq 0.83$.  The ensuing evolution is characterized by the
drift of the boundaries between active and empty regions.  A given realization
stops either when it attains the absorbing state, or when the number of
particles indicates that the active phase has filled the entire system
(we use a particle number of $N_{stop} = 0.84 L$ as the criterion for this event).
Figure \ref{qcnc7} shows a typical history for $L=2000$
and $n_0=1500$.  (The graph shows the mean density in blocks of 50 sites,
with time increasing downward, in steps of 10\,000 time units between each density profile.)
The boundaries between active and inactive regions appear to follow
independent, unbiased random walks.

\begin{figure}[ht]
\begin{center}
\epsfxsize=150mm
\epsffile{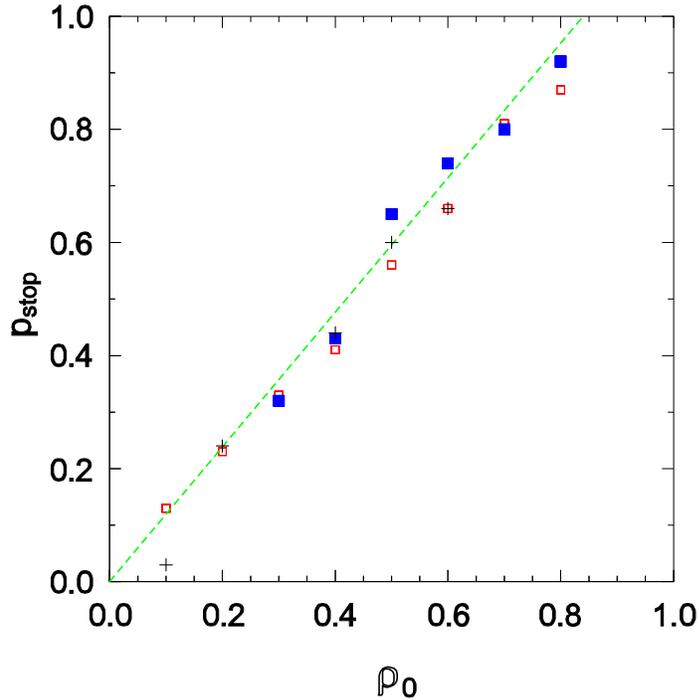}
\caption{(Color online) Probability $p_{stop}$ versus $\rho_0 = n_0/L$
in the TCM with an inhomogeneous initial
configuration.  $+$: $L=500$; open squares: $L=1000$; filled squares: $L=2000$.
Dashed line: $p_{stop} = n_0/N_{stop}$ as expected for a random walk.}
\label{tcrnc}
\end{center}
\end{figure}

If the boundaries can be represented by independent, unbiased random walkers, then the number $N(t)$ of
particles (which is approximated by $\rho_{QS}$ times the size of the active region), should also follow an
unbiased random walk.  The walk starts at $N=n_0$ and is subject to absorbing frontiers at $N=0$
and $N=N_{stop}$.  Well known results on random walks \cite{vankampen}
then imply that the probability $p_{stop}$
of reaching $N_{stop}$ before $N=0$ is given by $n_0/N_{stop}$.   We estimate $p_{stop}$ in sets of
100 realizations, on rings of 500, 1000, and 2000 sites; the linear trend evident in Fig. \ref{tcrnc}
supports the fluctuating boundary interpretation.  This in turn suggests that at high diffusion rates,
the dynamics of active and inactive domains is effectively that of CDP.

\begin{figure}[ht]
\begin{center}
\epsfxsize=150mm
\epsffile{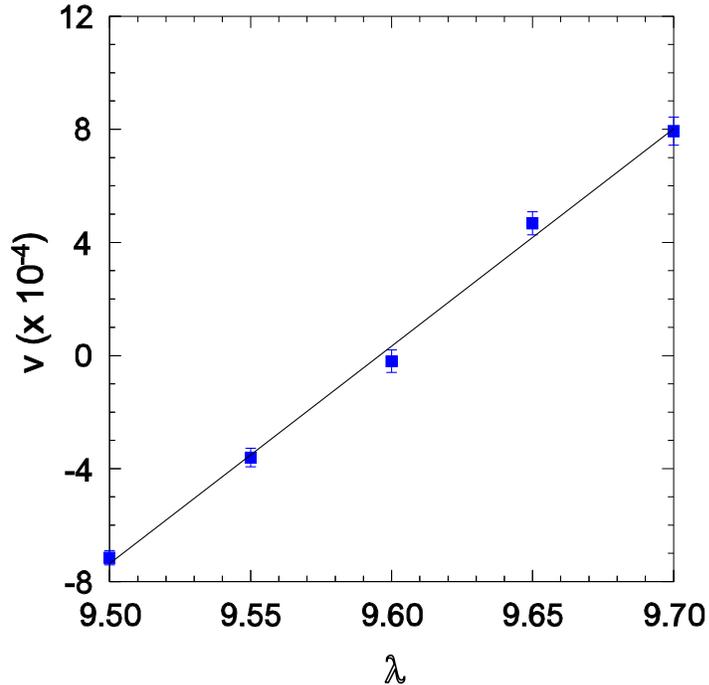}
\caption{(Color online) Interface drift velocity $v$ versus creation rate $\lambda$, for $D=0.98$,
$L=3000$.  The straight line is a least-squares linear fit to the data.}
\label{wv3k}
\end{center}
\end{figure}

Away from the phase transition, we expect the interface to drift on the average, advancing
into the inactive region for $\lambda > \lambda_c$ and vice-versa.  We determined the mean
interface velocity for $D=0.98$ on rings of 3000 sites, using $n_0 = 1500$.  After allowing the system to
relax for 50$\,$000 MCS, we record the density profile $\rho_i$, and determine the interface
position $x_i$ via the criterion $\rho(x_i) = \rho_B/2$, with $\rho_B$ the bulk particle density
at the $\lambda$ value of interest.  We then allow the system to evolve for an additional 20$\,$000
MCS, and again determine the interface position.    Fig. \ref{wv3k} shows
the interface drift velocity $v$, as determined in samples of $3 \times 10^4$ realizations,
varying linearly with $\lambda - \lambda_c$; linear regression yields
$v = 0$ for $\lambda = 9.60(1)$, in agreement with other estimates of the transition point.
(We verified that the interface velocity
obtained using a ring of 5000 sites, and observation times of $10^5$ and $2 \times 10^5$ MCS,
agrees to within 2\% with the value obtained using the smaller system.)

\begin{figure}[ht]
\begin{center}
\epsfxsize=150mm
\epsffile{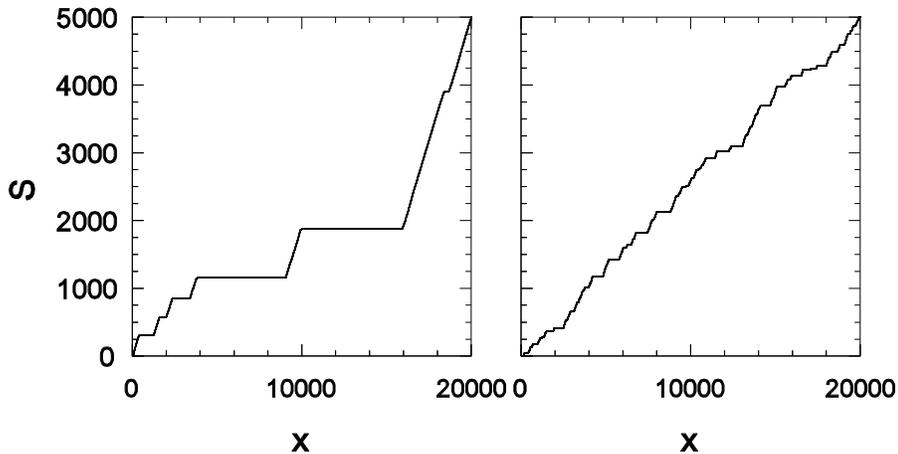}
\vspace{-3cm}

\caption{Cumulative particle number $S(x)$ versus position $x$ in a typical configuration. Left: $D=0.98$,
$\lambda=9.608$; right: $D=0$, $\lambda=12.015$.}
\label{sc2021}
\end{center}
\end{figure}

The above results show that for large $D$ the system is divided into well defined active and empty
regions.  A typical configuration of a large system ($2 \times 10^4$ sites) at the transition ($D=0.98$,
$\lambda=9.608$) bears this out.  In order to visualize the configuration of a large system, we plot
the cumulative particle number $S(x) = \sum_{i=1}^x \sigma_i$ versus position $x$, where $\sigma_i$
is an indicator variable taking values of 0 and 1 at empty and occupied sites, respectively.
Thus the local density $\rho(x)$ corresponds to the slope of the graph at $x$, with empty regions
corresponding to horizontal lines.  In these studies we initially occupy half the sites, randomly,
so that initially the local density is $\simeq 0.5$.  In the initial phase of the evolution, the
global density rapidly grows to about 0.8; thereafter it begins to fluctuate, as empty regions form.
The configuration shown in Fig. \ref{sc2021}, for a time of about $2.6 \times 10^8$ MCS (comparable
to the simulation times in \cite{park09}), consists of a series
active regions, with density $\rho \simeq 0.845$, and empty regions, giving an overall density of
0.25.  Of note is the high density in the
active regions, and the similarity of the density in active regions separated by large inactive
gaps.  We verified that for this choice of $D$
and $\lambda$, the density in active regions $\rho_a
= 0.840(5)$, {\it independent of system size and of overall density}.
The system reaches the absorbing state via fluctuations of the boundaries between
active and inactive regions, which eventually drive the active fraction to zero, while the
active region density remains constant.
For comparison, in Fig. \ref{sc2021}  we also show a typical
configuration for $D=0$ and $\lambda=\lambda_c(0) = 12.015$.  The initial condition is the same
as for $D=0.98$, and the simulation is again halted when $\rho$ falls to 0.25, which occurs
at $t \simeq 17\,000$ MCS for these parameters.
In this case the empty regions are typically much smaller than under rapid diffusion,
and the local density in active regions varies considerably.

\subsection{Effect of a weak source}

We turn next to studies of the stationary order parameter in the presence of a
weak source, $h$, defined as the probability, per unit time and per vacant site,
to insert a particle.  In \cite{Dick1st} a weak source was used to demonstrate
hysteresis; here, we consider the scaling of the order parameter as $h \to 0$
at the transition.
For $D=0$, we determine $\rho$ at $\lambda = 12.015$, varying $h$ between
$10^{-9}$ and $10^{-5}$.  For each $h$ value, a series of lattice sizes
(from 5000 to 50\,000 sites) are used to estimate the limiting infinite-size value of
$\rho$.  We verify the scaling law $\rho \propto h^{1/\delta_h}$, with $1/\delta_h = 0.109(1)$,
in agreement with the value expected for directed percolation in one spatial dimension,
$1/\delta_h = 0.10825(3)$ \cite{DickMar}.

\begin{figure}[ht]
\begin{center}
\epsfxsize=120mm
\epsffile{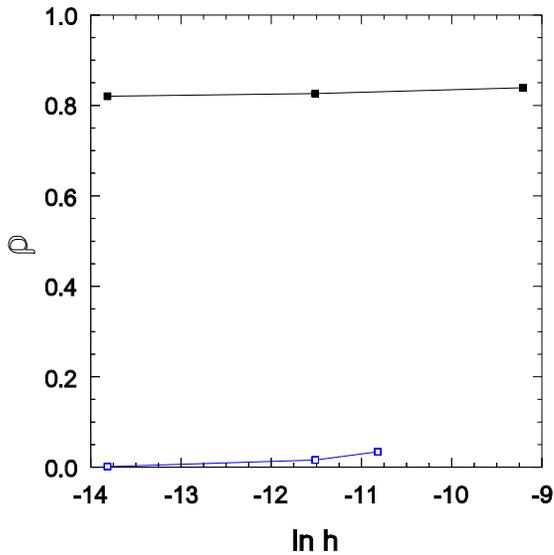}
\caption{(Order parameter versus source intensity $h$ for $D=0.98$, $\lambda=9.60$, $L=1000$.
Upper curve: initial density unity; lower: initial density zero.}
\label{tcrsd98}
\end{center}
\end{figure}

For $D=0.98$ and $\lambda = 9.60$, we observe a very different scenario.  For a given value
of $h$, two values of $\rho$ are found, depending on the initial density.  For large initial densities,
$\rho$ approaches a value of about 0.815 as $h \to 0$, while for a low initial density, $\rho \propto h$
(see Fig.~\ref{tcrsd98}).  The order parameter, moreover, is essentially independent of system size for
$L \ge 1000$.  These results are consistent with a discontinuous transition, and the absence of
critical scaling, for $D=0.98$.

\subsection{Quasistationary order parameter}

\begin{figure}[ht]
\begin{center}
\epsfxsize=140mm
\epsffile{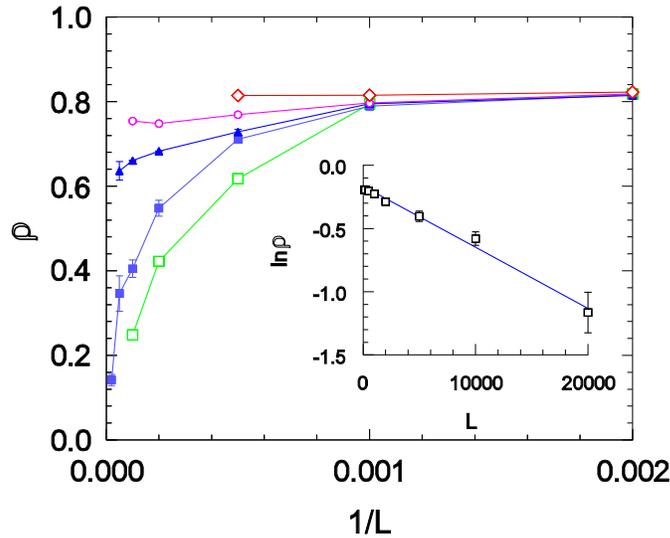}
\caption{QS order parameter $\rho$ versus reciprocal
system size for $D=0.98$ and (lower to upper)
$\lambda$ = 9.60, 9.605, 9.61, 9.62, and 9.65.  Inset: semi-log plot of $\rho$
versus $L$ for $\lambda=9.6084$.}
\label{tcfs}
\end{center}
\end{figure}

To close this section we report results on the QS value of the order parameter as a function of system size.
In Ref. \cite{MD07}, estimates for the limiting ($L \to \infty$) value of $\rho_{QS}$ were found to
exhibit a discontinuity at the transition, for $D=0.95$; here we focus on $D=0.98$.
We determined $\rho$ in QS simulations of duration $t_{max} = 10^7$ MCS
(for $L=100$) up to $t_{max} = 2 \times 10^9$ MCS
(for $L=10^4$), allowing the first 10\% of the time for relaxation.
In Fig. \ref{tcfs}
we plot the QS order parameter versus $1/L$, for $\lambda$ values near the transition.  The curves divide
into two families. One set (for $\lambda \leq 9.605$) approaches zero as $L \to \infty$, while for larger
values of $\lambda$, the density approaches a nonzero limiting value.
In the minute interval $9.605 < \lambda < 9.610$ the
limiting ($L \to \infty$) value of the order parameter $\rho$ jumps from zero to about 0.6.  The inset shows
that at the transition point, $\lambda = 9.6084$, $\rho$ decays exponentially with system size.
At a continuous transition one expects the density to decay as a power-law, $\rho_{QS} \sim
L^{-\beta/\nu_\perp}$.

\section{Why the transition is continuous, and why it appears to be discontinuous}

At various points in this discussion we have drawn an analogy between the TCM at high diffusion rates
and compact directed percolation.  The picture that emerges from the simulations
reported above, and from some of the earlier studies \cite{fiore,CF06-1st},
is that in this regime the system contains essentially two kinds of regions, one of high density, the other
empty.  The boundaries between these regions perform independent random walks,
leading eventually to extinction of activity.

The above scenario corresponds to the phase transition in CDP, and suggests that we construct
a reduced description in which blocks of $\ell$ sites in the active region (with particle density on the
order of $\rho_a \approx 0.85$) correspond to sites in state 1 of the CDP, and blocks of $\ell$ empty sites
correspond to sites in state 0 in the CDP.  The dynamics of the CDP consists exclusively of
random walks performed by the interfaces between strings of 0s and strings of 1s.
As we vary $\lambda$ through its critical value in the TCM, the drift velocity of the
interfaces in the corresponding CDP passes through zero, and the asymptotic density of 1s jumps
from zero to one.

If the above caricature of the TCM as an effective compact directed percolation model
were valid, a discontinuous transition would be guaranteed.  There are, however, two additional
processes that must be taken into account.  Evidently, gaps (strings of empty sites) can arise
within active regions, else the TCM starting from a fully occupied lattice would never reach
the absorbing state.  A fundamental point is that the process of gap nucleation, while essential
to the TCM dynamics, occurs at an extremely small rate.  By ``gap nucleation" we mean the generation
of a gap large enough (of $g^*$ sites, say) that its boundaries fluctuate independently of one another.
Gaps of size $g \geq g^*$ are equally likely to grow or to shrink, whereas
smaller gaps tend to shrink, due to particles diffusing
in from the adjacent occupied regions.

\begin{figure}[ht]
\begin{center}
\epsfxsize=140mm
\epsffile{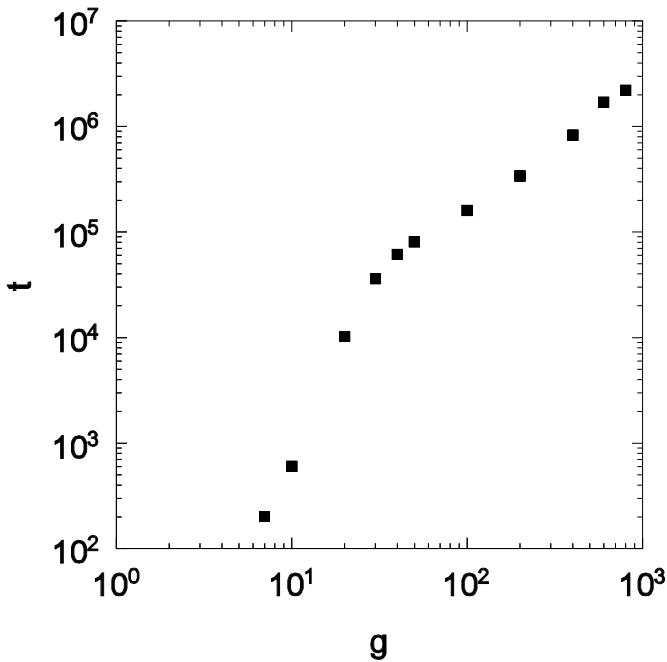}
\caption{Mean first-passage time $t$ for appearance of a gap of size $g$. Parameters
$L=10^4$, $D=0.98$, $\lambda=9.608$.
}
\label{tgap}
\end{center}
\end{figure}

The data of Fig. 3 permit an order of magnitude estimate
of the rate of gap nucleation: the density begins to fall appreciably from its plateau value
at $t \approx 10^5$, and there are ${\cal O} (10^5)$ sites in the system, giving a rate of
$\kappa \sim 10^{-10}$ per site.  In Fig. \ref{tgap} we plot the mean first-passage time $t$ for the
appearance of a gap of size $g$ in a system of $10^4$ sites (parameters $D=0.98$ and $\lambda=9.608$),
starting with all sites occupied.  (In these studies $t$ is estimated using samples of $N_r= $ 50 to 500
realizations, with smaller $N_r$ for larger system sizes.)
The first-passage time grows rapidly for smaller sizes and then crosses over to a slower growth
around $g \simeq 30$, at which point
$t \sim 6 \times 10^4$.  Identifying this crossover size as
$g^*$ gives a nucleation rate of $\kappa \sim 10^{-9}$ per site.

It is natural to take the block size $\ell$ in the CDP mapping as the critical gap size $g^*$,
so that a one-site gap in the equivalent CDP process is equally likely to grow or to shrink to zero.
Since we map $g^*$ TCM sites to a single site in the CDP,
the effective gap nucleation rate in the latter is then $\kappa_{eff} = g^* \kappa$.
The estimates for the nucleation rate and for $g^*$ given above yield  $\kappa_{eff}$ in the
range $10^{-9} - 10^{-7}$ per CDP site.

As noted above, a small gap can shrink when particles diffuse in from outside.  Can a large gap
be destroyed in this manner?  To answer this, consider an interface (treated as fixed, for the sake
of this argument) between large active and empty regions: the local density $\rho(x) = \rho_a$
for $x<0$ and $\rho \simeq 0$ for $x>0$.  Think of the edge of an active region as a particle source.
Particles are emitted at a rate of order $s = 1-D$, diffuse away at rate D, and
decay at rate $\gamma = (1-D)/(1+\lambda)$.  In the stationary state, a continuum diffusion analysis
yields a local density for $x>0$ of

\begin{equation}
\rho(x) = s \int_0^\infty dt e^{-\gamma t} \frac{e^{- x^2/4Dt}}{\sqrt{4 \pi D t}}
\end{equation}
\vspace{.2em}

\noindent giving $\rho(x) \simeq s e^{-x/w}$ with an interface width $w = \sqrt{D/\gamma}$.
For $D=0.98$ and $\lambda=9.60$ this gives $w \simeq 23$; for these parameters simulation shows
that near the edge
of a large gap, the density decays to zero $\propto e^{-x/25}$.  Nucleation of an active region
inside a gap occurs at a rate $\approx (1-D) [\rho(x)]^3$, making the probability
of nucleating activity deep within a gap negligible. For $x=100$ and $D$ and $\lambda$ as above,
for example, we find a nucleation rate of $\sim 10^{-13}$.
The essential point is
that $\rho(x)$ decays exponentially, so that nucleation of activity is limited to the
neighborhood of the edges, whereas the nucleation of gaps can occur anywhere
inside an active region.

Thus we conclude that the TCM at a high diffusion rate is equivalent to compact directed percolation
with a very small, but nonzero rate of gap formation within clusters of 1s.  But this process
is in turn equivalent (insofar as scaling properties are concerned) to the Domany-Kinzel
cellular automaton (DKCA) \cite{domanykinzel} with $p_1 \simeq 1/2$ and $p_2 = 1 -\kappa_{eff}$.
(Recall that in the one-dimensional DKCA, $p_1 $ is the probability of a site taking state 1, given
one neighbor in state 1, and one in state 0, at the preceding time, and that $p_2$ is the probability
of state 1 given that both neighbors are in state 1 at the previous step.)  CDP corresponds to the
line $p_2 = 1$.  From the work of Janssen \cite{janssenDK} and of L\"ubeck \cite{lubeckDK}
we know that for $p_2=1-\epsilon$, the transition occurs at $p_c = 1/2 + {\cal O}(\sqrt{\epsilon})$.

The phase transition of the DKCA is discontinuous {\it only} for $p_2=1$;
for {\it any} $p_2 < 1$, it is continuous and
belongs to the DP universality class.  We are led to the same conclusion regarding the TCM: for any
$D < 1$, there is a small but finite rate of nucleating gaps within active regions, so that the
effective value of $p_2$ is slightly less than unity.

Given the nearness of the equivalent DKCA to the line $p_2=1$, it is not surprising that simulations
of the TCM using lattice sizes $L$ and simulation times $t_m$ yield an apparently discontinuous
phase transition, for $Lt_m < 1/\kappa$.  It is only for large systems and long simulation times
that the effects of gap nucleation become apparent, as in the studies of \cite{Hayefirsto}
and \cite{park09}.  (Indeed, the results shown in Fig. 3 are also compatible with DP-like
decay of the order parameter.)  But numerical studies of the stationary order parameter
can be expected to show discontinuous behavior, this near the CDP line.  For similar reasons,
it is not surprising that $n$-site approximations, using clusters of fewer than twenty sites,
miss the effect of gap nucleation; much larger clusters would be needed to capture this properly.

It is perhaps worth recalling that CDP-like spreading behavior is observed in a {\it surface modified}
version of directed percolation in one spatial dimension \cite{mendes96}.  In this case, propagation
of activity at the edges of the active region occurs with a different creation rate ($\lambda'$, say),
than in the bulk, which has a creation rate of $\lambda_B$.  Let $\lambda_c$ denote the critical creation rate for
the original problem, that is, for $\lambda' = \lambda_B$.
For $\lambda' < \lambda_c$, the phase transition occurs at some bulk creation rate $\lambda_B > \lambda_c$,
which means that the bulk has a finite activity density even for $t \to \infty$.  Thus the active cluster
is compact, and the scaling behavior is that of CDP.  The essential difference between this model
and the TCM is that in the former case large gaps cannot be nucleated within the active region: since
$\lambda_B > \lambda_c$ at the transition, gaps tend to shrink.

\section{Discussion}

We have presented various pieces of evidence suggesting that the order parameter is a discontinuous
function of the creation rate
in the triplet creation model at high diffusion rates (our numerical
studies focus on $D=0.98$).
We do not find evidence of stable coexisting active and
inactive regions; the boundaries between these regions are observed to fluctuate,
as asserted in Ref. \cite{Hayefirsto}.
The numerical evidence in favor of a discontinuous transition includes hysteresis (under a weak source),
the initial-density dependence of the order parameter at later times (with or without a
particle source), the high particle density within active regions, the random-walk-like
fluctuations of the boundaries between active and inactive regions, and the apparent absence of
power-law scaling of the QS order parameter as a function of system size, at the transition point.

Despite the numerical evidence in favor of a discontinuous transition,
a mapping of the TCM to a effective dynamics resembling that of
compact directed percolation, but with a very small gap nucleation rate, leads to the conclusion
that the transition is continuous for any $D<1$.  The studies reported above, suggesting a discontinuous
transition, were performed using relatively small systems and/or limited times.  For example, to see
the effect of gap nucleation in the studies with a weak source of activity, the source strength
$h$ would have to be much smaller than the gap nucleation rate $\kappa$.

The failure of $n$-site approximations to give a clear indication of the nature of the transition
may again be attributed to the very small gap nucleation rate for $D \simeq 1$.  These approximations,
however, are problematic even for $D=0$: the values predicted for $\lambda_c$ do not converge
monotonically to the correct value with increasing $n$ \cite{ff09}.

The continuous nature of the transition in the one-dimensional TCM
was of course asserted some time
ago by Hinrichsen \cite{Hayefirsto}, and received further support in Park's simulations \cite{park09}.
Our argument nevertheless contributes to an intuitive understanding of this result, and might
provide the basis for a rigorous demonstration of the continuous nature of the transition.
\vspace{1.5em}

\noindent {\bf Acknowledgments:}

We are grateful to Jos\'e F. Fontanari and M\'ario J. de Oliveira for helpful discussions.
Support from Hungarian research fund OTKA (Grant Nos. T046129, T77629)
is acknowledged.
The authors acknowledge access to the HUNGRID, Clustergrid and the
supercomputer of NIIF Budapest.  RD acknowledges support from CNPq,
and Fapemig, Brazil.

\end{document}